# How accurately defined are the overtone coefficients in the Gd(III)-Gd(III) RIDME?

Mykhailo Azarkh[a *], Katharina Keller[b], Mian Qi[c], Adelheid Godt[c], Maxim Yulikov[b *]

[a]Department of Chemistry, University of Konstanz, Universitätsstraße 10, 78457 Konstanz, Germany

[b]Laboratory of Physical Chemistry, Department of Chemistry and Applied Biosciences, ETH Zurich, Vladimir-Prelog-Weg 2, 8093 Zurich, Switzerland

[c]Faculty of Chemistry and Center for Molecular Materials (CM2), Bielefeld University, Universitätsstraße 25, 33615 Bielefeld, Germany

To whom the correspondence should be addressed:

mykhailo.azarkh@uni-konstanz.de (M.A.)

maxim.yulikov@phys.chem.ethz.ch (M.Y.)

*Abstract*

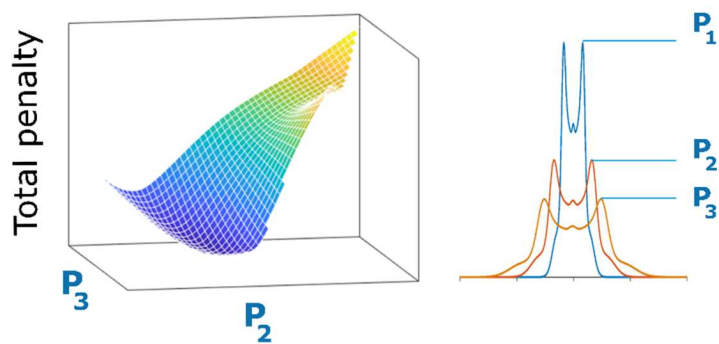

Relaxation-induced dipolar modulation enhancement (RIDME) is a pulse EPR technique that is particularly suitable to determine distances between paramagnetic centers with a broad EPR spectrum, e.g. metal-ion-based ones. As far as high-spin systems (S > ½) are concerned, the RIDME experiment provides not only the basic dipolar frequency but also its overtones, which complicates the determination of interspin distances. An r.m.s.d.-based approach for the calibration of the overtone coefficients is proposed and illustrated for a series of molecular rulers doubly labelled with Gd(III)-PyMTA tags. The constructed 2D total-penalty diagrams clearly show that there is no unique set but rather a certain pool of overtone coefficients, which can be used to extract distance distributions between high-spin paramagnetic centers as determined from the RIDME experiment.

*Introduction*

Pulse dipolar spectroscopy (PDS) in EPR offers a powerful set of techniques to determine distances between unpaired electrons[1,2]. When combined with site-directed spin labeling[3-5], PDS becomes a valuable tool in structural biology, allowing to determine structure of biomacromolecules[6,7] or biomolecular complexes[8] and to follow structural changes they undergo[9]. Depending on the number of subunits in a biomolecule or biomolecular complex under study, as well as on the sample preparation conditions, a certain type of spin label or a combination of different types of spin labels can be used[10-12]. Besides the nitroxide-based spin labels, substantial attention has been paid in the last few years to optimize PDS approaches for paramagnetic metal-ion-based spin labels[13-18]. In particular, Gd(III)-based spin labels attract significant attention, since they exhibit favorable relaxation properties and stability against intracellular reduction[19-23]. Relaxation-induced dipolar modulation enhancement (RIDME) is one of the possible PDS experiments for distance determination with spin pairs that include Gd(III) ions[24-26]. Given the fact that a spectrum of Gd(III) spreads over up to 2 GHz, a particular advantage of RIDME is that for all excited observer spins, the coupling to partner spins is detected regardless of the resonance frequency of the latter, without requiring large resonator bandwidth or excitation bandwidth of the microwave pulses.

Here we take an example of the Gd(III)-Gd(III) RIDME experiment, for which the largest set of experimental data and analysis was published up to date[27-30]. Since, in the Gd(III)-Gd(III) RIDME experiment, the modulation due to the dipolar spin-spin coupling is induced by spontaneous spin flips, dipolar frequency overtones (DFOs) are excited in addition to the main dipolar frequency. Their relative contributions are determined by the probabilities of changing the electron spin projection by more than one quantum in a single spin flip or in multiple spin flips. This holds for other types of high-spin centers as well. The presence of such overtones complicates data processing. DFOs, if not accounted properly during the data processing, lead to artefacts in the distance distribution. Consequently, for distance determination by Gd(III)-Gd(III) RIDME, the fractions of DFOs have to be taken into account. So far the most detailed calibration of DFO coefficients (DFOCs) was performed on a series of Gd-PyMTA-based spectroscopic rulers[27,30] and revealed several important points: (i) Only the primary dipolar frequency and its double and triple harmonics need to be taken into account, while the contributions from higher harmonics are negligible; (ii) DFOCs appear nearly perfectly constant within the RIDME mixing time, and almost constant within the practically relevant temperature range; (iii) DFOCs are distance-dependent for inter-spin distances below 3.0 nm, while they become approximately distance-independent for distances above 3.0 nm; (iv) The RIDME data fitting is rather sensitive to the proper selection of the ratio between the weight of the primary dipolar frequency and the sum of the weights of the second and third DFO, while the quality of the data fit is only weakly dependent on the small variations of the two DFOCs as long as their sum is kept constant; (v) For the quality of the DFO calibration, it is important to use reference samples with narrow distance distributions, so that artefact distance peaks due to DFOs are resolved. Please note that these findings are specifically for the Gd-PyMTA-based rulers, and should be transferred to other Gd(III) complexes with care.

Furthermore, the negligibly weak contributions from higher overtones point to a transient balance rather than an established thermodynamic equilibrium of the spin states of the partner spins. The approximate balance of DFOCs is further revealed by the observation that below a certain threshold distance, which was estimated for the Gd-PyMTA complex to be somewhere around 3.0 nm, DFOCs start to be distance dependent. Thus, when we consider the distance analysis above this threshold, then establishing a fixed set of DFOCs means that we find one set of such coefficients that can provide data in the whole distance range with a quality sufficient for bio-EPR applications. This, however, does not mean that this set of DFOCs is exact and unique. Essentially, this procedure consists of determining the best set of DFOCs for each reference sample and then establishing an "average set" of DFOCs that would result in the minimal error over the entire calibration set. While the last step is methodologically clear and depends in an unambiguous way on the type of the penalty function taken for the error determination, the step of determining this penalty function and the best set of DFOCs for each reference sample potentially allows for different solutions.

Slightly different sets of DFOCs appear as a result of different background corrections, variation in noise level and the presence of distance distribution artifacts that are not due to the DFOs. For instance, it has been proposed recently that the DFOCs for Gd-PyMTA-based rulers are slightly different in $D_2O$ and $H_2O$[30]. While this difference might in principle point to the change of the DFOCs upon the change of the spin label environment (protonated vs. deuterated glassy solvent), the differences in the fit quality between the two mentioned sets of the DFOCs are not particularly large, which implies that the differences in the values of the DFOCs might also be a result of differences in the calibration procedures. Since we expect that the topic of calibrating DFOCs for different types of Gd(III) complexes will be of importance in the near future, we consider it important to re-discuss the calibration procedure of the DFOCs in detail and, most importantly, determine the accuracy limitations of this procedure.

Here, we present a detailed step-by-step discussion of a DFOCs calibration procedure for Gd(III)-Gd(III) RIDME experiment performed for a series of Gd-PyMTA-based rulers. This allows us to compute 2D total-penalty diagrams nicely visualizing the area within which DFOCs can be potentially selected. The paper is organized as follows. First, we discuss the key steps in determining the best set of DFOCs for one particular sample. This includes the analysis of the background correction procedure, the form factor fit quality, and the artefact distance peaks intensities. Second, we discuss the determined range of possible solutions. Finally, we summarize the output of the analogous detailed analysis for other Gd-PyMTA-based rulers and give a comparison of the DFOCs calibration of one Gd-PyMTA-based ruler in protonated and deuterated matrix.

## *Experimental details and data analysis procedures*

**Gd-PyMTA-based rulers.** Stiff molecular rulers that bear two Gd-PyMTA complexes connected by a linear shape-persistent linker were used[27,31,32]. The distances between the Gd(III) ions in these rulers are 2.1, 3.0, 3.4, 4.3, 4.7, and 6.0 nm[27,30]. RIDME samples were prepared by dissolving the rulers in either $D_2O$ or $H_2O$ (20-50% by volume glycerol-$d_6$ or glycerol, respectively) to yield a final ruler concentration in the range of 25–300 μM[27,30].

**RIDME measurements.** RIDME measurements were performed at Q- and/or W-band, as reported recently. The details are given elsewhere[27,30]. For each interspin distance one set of data was selected. The corresponding temperature and mixing times are given in Table 1.

Table 1. Selected experimental parameters for the datasets used

|                | 2.1 nm | 3.0 nm (prot.) | 3.0 nm (deut.) | 3.4 nm | 4.3 nm | 4.7 nm | 6.0 nm |
|----------------|--------|----------------|----------------|--------|--------|--------|--------|
| Mixing time, µs | 12    | 8              | 12             | 10     | 24     | 24     | 16     |
| Temperature, K | 20     | 10             | 10             | 20     | 10     | 10     | 10     |
| MW band        | W      | Q              | W              | W      | W      | W      | W      |

**Data analysis.** Data analysis was performed with Matlab. Built-in functions from OvertoneAnalysis were used for processing of the experimental data, background correction, and Tikhonov regularization.

Background correction. A background model was chosen by selecting the background start of the RIDME time trace and fitting a stretched exponential to it. The penalty functions for the background corrections were defined as r.m.s.d. values for the fits of the form factor and the Pake pattern, which were obtained after processing the background corrected data with Tikhonov regularization disregarding DFOs, i.e. $\{P_1 = 1, P_2 = P_3 = 0\}$.

Fit quality as a function of DFOCs. The form factor obtained after the background correction was processed by Tikhonov regularization, while the kernel function contained a pre-defined set of DFOCs $\{P_1, P_2, P_3\}$. The two form factor penalty functions related to this set of DFOCs were the r.m.s.d. values for the fits of the form factor and the Pake pattern. These two r.m.s.d. values were combined into a total penalty function, as described in the Results and Discussion section.

Artefacts in the distance distribution. DFOCs, when under- or overestimated, produce artefacts in the distance distribution below or above the expected distance, respectively. The distance penalty functions with respect to these distance artefacts were computed as follows. A distance distribution obtained by Tikhonov regularization as described above was normalized to the intensity of the expected distance (i.e. at 2.1, 3.0, 3.4, 4.3, 4.7 and 6.0 nm, according to the molecular rulers used) and cut into two parts at this distance. The distance probabilities were summed up for each sub-distribution separately; each of the two resulting numerical values contained the contributions from the artefact distances and a constant contribution from the real distances. From this procedure, two distance penalty functions for a given set of DFOCs are obtained. One for the distance artefacts below the expected distance, in the following lower-distance penalty function, and the second for the distance artefacts above the expected distance, in the following upper-distance penalty function.

Error bars for DFOCs. The error bars are calculated from the total-penalty diagram. First, the minimum total penalty is found. Second, an area on the total-penalty diagram is determined, which corresponds to 10% deviation from the minimum total penalty. The DFOCs are read out from this area. For each DFOC, a minimum and maximum value is read out and assigned to the lower and upper limit for the error bar, respectively.

## Results and Discussion.

**General approach to calibrate DFOCs.**

The workflow for calibrating DFOCs is given in Fig. 1. A background corrected form factor is processed for all combinations of DFOCs, which are systematically varied, and for each combination of DFOCs the distance distribution is obtained by Tikhonov regularization. This set of distance distributions and the corresponding form factor fits is the subject of further analysis.

The quality of the form factor fit – both in time and frequency domain – is determined as a function of DFOCs (Fig. 1a, b). Next, the space of DFOC combinations is restricted to those that produce no artefacts in the distance distribution above the expected distance based on the upper-distance penalty function (Fig. 1c). For this restricted set of DFOC combinations, the fit quality criterion is multiplied with the lower-distance penalty function, i.e. the distance penalty function due to distance artefacts below the expected distance (Fig. 1d, e). In the last step, both penalty functions are combined by multiplication to provide the total penalty function for the final analysis (Fig 1f). The best set(s) of DFOCs corresponds to the minimum of this total penalty function.

We should first discuss in detail the key data analysis steps in the RIDME experiment, and give some estimates in the underlying uncertainties.

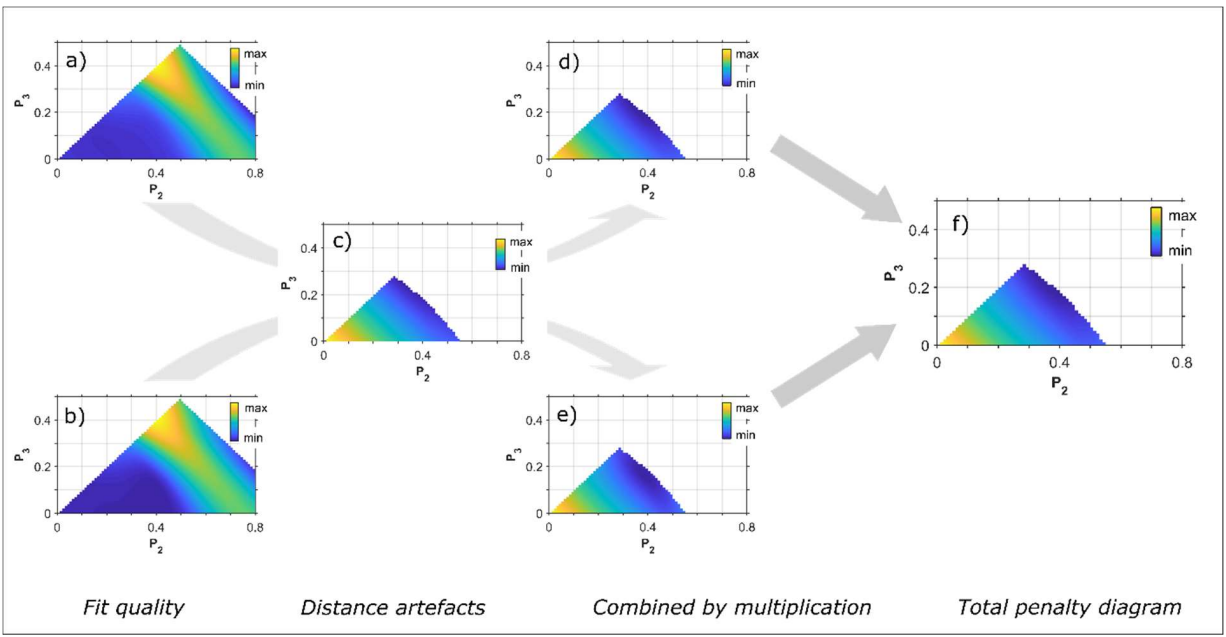

**Figure 1.** A schematic workflow for calibrating DFOCs, illustrated for the Gd-PyMTA-based ruler with 4.3 nm interspin distance. (a) and (b) The r.m.s.d. plots for the form factor and the Pake pattern fit, respectively. (c) The space of DFOCs that do not produce distance artefacts above the expected distance; the lower-distance penalty function is given in color. (d) and (e) The combined penalty function due to the fit quality and the distance artefacts below the expected distance (lower-distance penalty function) for the form factor and the Pake pattern, respectively. (f) The total penalty function.

**Step 1: Background correction.**

To a good approximation, the intermolecular background in the RIDME experiment can be described as a combination of two distributions, the first one formed by mono-exponential decay curves with different decay constants, and the second one formed by a distribution of different Gaussian decay curves[28]. In practice, RIDME background decay can be often fitted by a stretched exponential curve, provided the signal does not decay too close to zero. RIDME background curves decaying down to approximately 20% of their initial intensity were sufficiently well fitted by one stretched exponential curve, while the tail of the background decay trace was demonstrating stronger deviations from this law, so that fitting of the entire background trace required a sum of two stretched exponential functions[28]. Thus, for the best stability of the background fit, one should restrict the measured time range to the region which is still well fitted by a single stretched exponential curve. For this reason, the sensitivity advantage of RIDME comes at the cost of a smaller accessible distance range.

The most common first step in the analysis of PDS data is the fitting of background, followed by the determination of the form factor trace, although joint fitting of the background and form factor has been proposed as well[33,34]. Depending on the selected background model, somewhat different form factor traces are obtained, which then result in somewhat different best sets of DFOCs. The difficulty at this step appears because the best set of DFOCs are only obtained at the end of the calibration procedure, while for the presented formalism one needs to decide on the background model at this initial step. To uncouple these two calculation steps, we performed a fitting of the resulting form factor traces that included only the contribution from the primary dipolar frequency, without any overtones. This, of course, resulted in an unrealistic multi-peak distance distribution. However, such a fit produced the best possible match to the experimental form factor trace that still conforms to the non-negativity restriction for the distance distribution. The r.m.s.d. of the best form factor fit in this case has been taken as an estimate for the uncertainty in fitting DFOCs at the later steps in our procedure. Indeed, since all the models used are only approximately matching experimental data, the true DFOC values might deviate from the best fitted ones, and due to the inexact model, the correct DFOCs might produce a form factor fit that has a somewhat worse r.m.s.d. within the given model.

In principle, at this point one should have constructed a set of background fits, for which the form factor fitting with only primary frequency contributions produces r.m.s.d. values not exceeding the best one by more than a factor of two or so, as any of such fits would be approximately within the accuracy limits of the used model. We have omitted such a set of background fits for two reasons. First, the uncertainty appearing at the step of actual calibration of DFOCs substantially exceeds the uncertainties due to different background correction (vide infra). Second, including such a set of different background models into the DFOCs fitting procedure would make the computational effort unrealistic. Thus, in the following procedure, we continue only with the form factors resulting from the background fit with lowest r.m.s.d. value.

**Step 2: Quality of the form factor fit.**

In the next step, we take the background-corrected form factor trace and fit it with all possible combinations of the three DFOCs ($P_1$, $P_2$ and $P_3$). Due to the very different distance dependence of the shape of the form factor function in the frequency and time domain, we consider both of them. Since only two out of three DFOCs are independent, due to the normalization $P_1+P_2+P_3=1$, this results in a 2D r.m.s.d. plots, shown in the Fig. 1a and 1b. In principle, if we only consider the match between experimental and computed form factor trace, the best fit must be given by the DFOC-free combination $\{P_1 = 1, P_2 = P_3 = 0\}$. Note also that more local minima are seen in these 2D plots. In particular, the 'RIDME valley' around $P_2 = 0.3$ and $P_3 = 0.1$ is visible in the frequency-domain 2D r.m.s.d. plot. (Fig. 1b) Depending on the choice of the regularization parameter in the Tikhonov regularization procedure, it might happen that the r.m.s.d. of the optimal solution in the RIDME valley is even better than that of the DFOC-free solution $\{P_1 = 1, P_2 = P_3 = 0\}$. Here, for simplicity, we have always taken the by-default suggested value of the regularization parameter. This calculation demonstrates that, on the one hand, it is possible to roughly guess correct DFOCs for the given system by computing the form factor time-domain and frequency-domain 2D r.m.s.d. plots, but, on the other hand, this calculation is not enough to unambiguously determine the best DFOCs. In particular, we must know the shape of the underlying distance distribution for selecting the correct RIDME valley.

**Step 3: Intensities of the artefact distance peaks.**

Proper calibration of DFOCs requires the minimal deviation of the computed distance distribution from the anticipated one. We note that for DFOCs $P_2$ and $P_3$ smaller than the best-solution ones, the artefacts in the distance distribution are mainly placed at distances shorter than the true distance peak. As one or both of these DFOCs increase above the best solution values, stronger artifacts at the distances above the true distance peak appear[27]. However, also artifact peaks caused by other effects, such as background correction or insufficient ESEEM averaging, may be present. The position of such peaks would also be influenced by the DFOCs. In the presented approach, we assume that the contribution of such additional artifacts is weak and that the best solution corresponds to the minimum probability to find a distance outside of the true distance peak. This criterion can be formalized in somewhat different ways, but, clearly, its usability relies on the fact that the true distance peak is narrow enough so that the artifact peaks are well resolved in the computed distance distribution. Here, we scaled all distance distributions, computed for different DFOC sets, for the same intensity of the true distance peak. We identified all overestimated DFOC sets from the upper-distance penalty function, i.e. the sum of distance probabilities of the subdistribution above the true distance maximum. The threshold is set to 10% with respect to the minimum of this upper-distance penalty function for all DFOC sets. Consequently, all DFOC sets that produce distance distributions with an upper-distance penalty function above the threshold are excluded at this point of the calibration procedure (Fig. 1c). For the remaining DFOC sets, the lower-distance penalty function, i.e., the sum of distance intensities for the subdistribution below the true distance maximum, is multiplied with each of the two form factor penalty function to obtain a combined penalty function.

**Calibrated set of DFOCs.**

The best-solution set of DFOCs is then found from a total penalty function – a product of the two combined penalty functions pertaining to the form factor fits in time and frequency domain and the artefact distance peaks. A result of such analysis is shown in Fig. 1f. The requirement of the absence of artifact peaks helps to properly place the minimum of the total penalty function. Importantly, as has been already discussed qualitatively in the earlier report[27], a correlated change of $P_2$ and $P_3$ coefficients, which does not modify the sum $P_2+P_3$, has only a very weak influence on the value of the penalty function. In other words, this calibration procedure can relatively well restrain the values for $P_1$ and the sum $P_2+P_3$, but it leaves more uncertainty for the particular values of $P_2$ and $P_3$.

**Calibration of DFOCs for Gd(III)-PyMTA-based rulers**

*A series of molecular rulers.*

The above described systematic r.m.s.d.-based analysis was conducted for several previously analyzed Gd-PyMTA-based rulers in the deuterated matrix $D_2O$. The 2D total-penalty function plots are shown in Fig. 2, and the results for the best DFOCs are summarized in Fig.3 and Table 2. In line with the previously discussed trend[27], Fig. 2 shows that the penalty function is only weakly changing for the correlated change of $P_2$ and $P_3$. The best-fit value for the coefficient $P_1$, and thus for the sum $P_2+P_3$, is nearly constant for the distances above 3 nm. The mean value of $P_1$=0.45 averaged over all distances above 3 nm is somewhat smaller than the previously proposed value of $P_1 = 0.51$[27]. However, first, these two values are essentially lying within the error bar of the calibration procedure. And second, the differences in these two calibrations are mainly due to the different treatment of the dipolar spectrum data as discussed in the following.

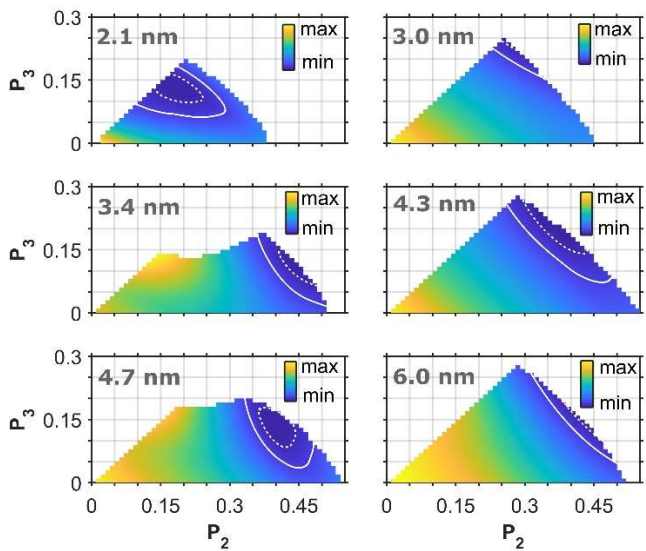

**Figure 2.** Analysis of the overtone coefficients for a series of Gd-PyMTA-based rulers in D$_2$O. The expected interspin distance is given in the figure. White contour lines indicate the area with 3% (dotted) and 10% (solid) deviation from the minimum of the total penalty function.

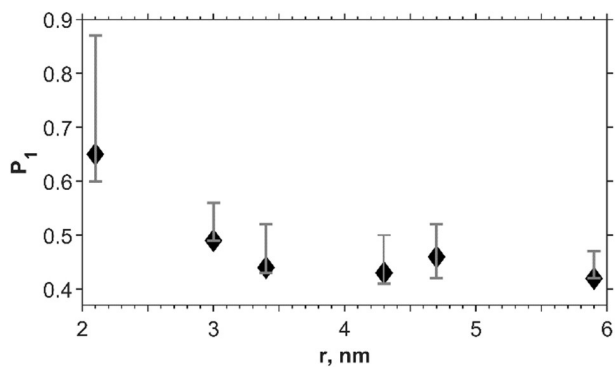

**Figure 3.** The P$_1$ coefficient as a function of the interspin distance. Error bars indicate the range for P$_1$ within the 10% area (cf. Fig.2).

Table 2. Overtone coefficients as a function of the interspin distance for the deuterated samples.

| Distance, nm | 2.1 | 3.0 | 3.4 | 4.3 | 4.7 | 6.0 |
|---|---|---|---|---|---|---|
| $P_1$ | 0.65 | 0.49 | 0.44 | 0.43 | 0.46 | 0.42 |
| $P_2$ | 0.22 | 0.29 | 0.42 | 0.37 | 0.40 | 0.43 |
| $P_3$ | 0.13 | 0.22 | 0.14 | 0.20 | 0.14 | 0.15 |

Another important point for such a formalized calibration approach can be seen in Fig. 4. The dipolar frequency pattern computed from the form factor trace reveals clear 'horn-like' features placed symmetrically with respect to zero frequency. The range of the DFOCs where these features are well reproduced by the fit is relatively narrow, which strongly helps to stabilize the DFOCs calibration procedure. In the time domain, due to the intrinsic properties of the Fourier transform, such deviations smear out over the entire time range, and thus the sensitivity of the r.m.s.d.-based fit to these deviations is very weak. However, in the frequency domain, the artifact due to the inaccuracies in the background correction is also 'condensed' near zero frequency. This increases the minimal r.m.s.d. of the fitted dipolar spectrum, and, thus, forces one to allow for a larger uncertainty of the fitted DFOCs. In Fig. 4 one can see a comparison between the best fit obtained with the fully formalized r.m.s.d.-based procedure described here and the previous manual calibration data, where the horn-like features in the dipolar spectrum were taken into account while disregarding the deviation in the zero-frequency region. It is clearly seen that, besides the zero-frequency region in the dipolar spectrum, the deviations from the experimental dipolar spectrum are somewhat smaller for the previously proposed DFOCs. Quite importantly, however, the here presented formal procedure is useful in estimating realistic uncertainties based on the r.m.s.d penalty function for the DFOCs calibration.

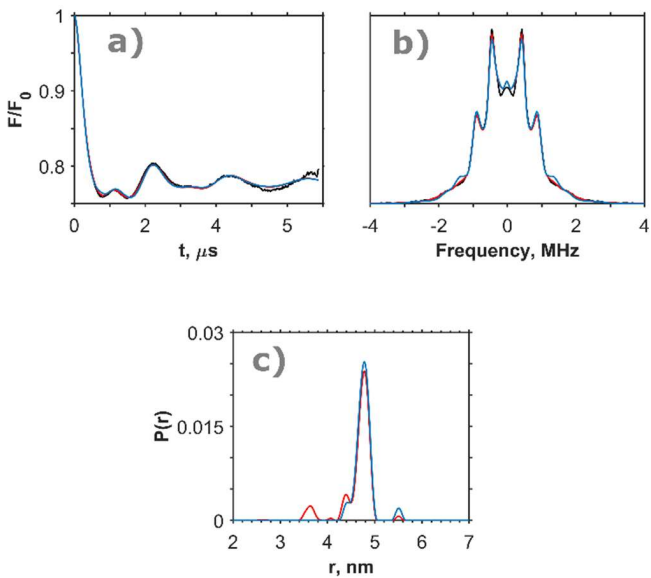

**Figure 4.** Comparative processing of the RIDME data for the Gd-PyMTA-based ruler with the expected distance of 4.7 nm: (a) the form factor, (b) the Pake pattern, (c) the distance distribution. Black – experimental data, colored – processed data with DFOCs from the 10% area [0.51 0.40 0.09] [27] (red) and from the averaged minimum of the total-penalty diagrams [0.44 0.41 0.15] (blue).

*Deuterated vs. protonated matrix.*

An example of the comparison of the DFOCs 2D total-penalty function plots for the cases of protonated and deuterated solvent are shown in Fig. 5. While the positions of the minima are not the same for these two plots, one can notice that realistic error bars of the DFOCs calibration exceed this difference. We can thus conclude that while it is not possible to state whether the DFOCs in these two samples are exactly the same, they are very similar, and for all practical purposes it is sufficient to use the same set of DFOCs for the cases of protonated and deuterated solvents.

Of course, here, the near identity of the DFOCs for protonated and deuterated samples is only shown for one type of Gd(III) complex. This should be, in principle, verified again, whenever a different type of Gd(III) complex is in use. Should this be a general rule for many such complexes, the DFOCs calibration procedure could be performed on the deuterated ruler solutions, and its results can be used also for the protonated solvent cases, e.g. in cells. Such a calibration procedure would require only about half the effort of calibrating the DFOCs separately for protonated and deuterated solutions.

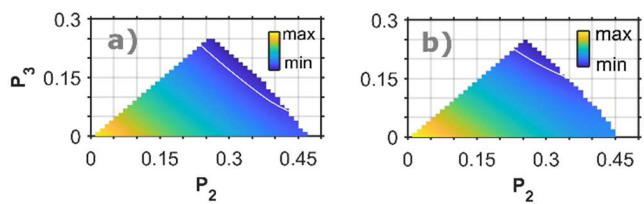

**Figure 5.** Total-penalty diagrams for a Gd-ruler with the expected interspin distance of 3.0 nm measured in (a) H₂O and (b) D₂O.

## *Conclusions*

We have presented a rather general procedure to estimate the mean values and trust regions of the DFOCs for high-spin RIDME experiments with Gd(III) complexes, which should perform the same also for the spectroscopically similar high-spin Mn(II) case (S=5/2). Our analysis confirms the previous conclusion that for spin-spin distances above 3 nm with the Gd(III)-PyMTA complex as spin label one set of DFOCs can be used for computing the distance distribution. Furthermore, we demonstrated that small differences in the previously reported DFOC values for protonated and deuterated samples are a matter of somewhat different choice of the best solution, but all so far determined DFOC sets fall into the relatively broad trust region of this calibration. Of course, our conclusions were made based on only one type of Gd(III) complex, and should be transferred to other types of Gd(III) complexes with care. Overall, this study confirms that the distance analysis in the Gd(III)-Gd(III) RIDME experiments can be performed in a rather simple way, as long as the distances below 3 nm are not present in the sample, while some additional calibration effort is required to include distances below 3 nm into the analysis. The cutoff distance of 3 nm still remains to be confirmed for other Gd(III) complexes as it may be influenced by the ZFS interaction. Fortunately, the calibration is only necessary e.g. in deuterated water/glycerol mixture, and can be afterwards used also for the protonated water/glycerol mixtures.

## *Acknowledgment:*

The work was supported by the Swiss National Science Foundation (grant 200020169057) and by the Deutsche Forschungsgemeinschaft (DFG) within SPP 1601 (GO555/6-2). We are grateful to Prof. Dr. G. Jeschke for reading the manuscript and for helpful discussions.